\documentclass[aps,showpacs,pra,twocolumn,amsmath,amssymb,superscriptaddress]{revtex4-1}
\usepackage{graphicx}% Include figure files
\usepackage{graphics}
\usepackage{subfigure}
\usepackage{bm,bbm}

\begin{document}

\title{A Quantum Theory of Cold Bosonic Atoms in Optical Lattices}

\author{Dagim Tilahun}
\affiliation{Department of Physics, The University of Texas at Austin, Austin, TX 78712}
\affiliation{Department of Physics, Texas State University, San Marcos, TX 78666}
\author{R. A. Duine}
\affiliation{Institute for Theoretical Physics, Utrecht University, Leuvenlaan 4, 3584 CE Utrecht, The Netherlands}
\author{A. H. MacDonald}
\affiliation{Department of Physics, The University of Texas at Austin, Austin, TX 78712}

\begin{abstract}
Ultracold atoms in optical lattices undergo a quantum phase transition from a superfluid to a Mott insulator as the lattice potential depth is
increased.  We describe an approximate theory of interacting bosons in optical lattices which provides a
qualitative description of
both superfluid and insulator states. The theory is based on a change of variables in which the boson coherent state amplitude is replaced by an
effective potential which promotes phase coherence between different number states on each lattice site.  It is illustrated here by
applying it to uniform and fully frustrated lattice cases, but is simple enough that it can easily be applied to spatially inhomogeneous lattice
systems.
\end{abstract}
\maketitle

\section{Introduction}

The observation \cite{GreinerNature} of a cold-atom quantum phase transition between superfluid (SF) and Mott insulator (MI) states was important on its own merits, and also 
because it suggested future experimental studies of clean highly controllable strongly correlated bosonic many-body systems.  The promise of early experiments has been borne out by studies 
that have demonstrated unprecedented experimental control in designing and investigating many body systems 
whose Hamiltonian's are known with a level of precision that 
is uncommon in condensed matter physics \cite{RevModPhys.80.885}. 
Cold atom systems are not, however, completely free of the {\em real world} complications that can confuse the
interpretation of experiments.  The most obvious troublesome complications in simulating
condensed-matter many-body physics problems using cold atoms are that experimental systems are always spatially inhomogeneous to some degree, and that they are often fairly small. 
In most cases, the spatial inhomogeneity is an undesirable consequence of an experimental necessity, 
for example the harmonic trapping potential employed in most cold atoms set-ups.
In some cases, though, it is the central focus of the experiment, 
as in studies of disorder in strongly correlated bosonic systems \cite{PhysRevLett.98.130404,DeMarco}.  In this paper we describe an approximate theory of strongly interacting bosons in an optical lattice that is sophisticated enough to achieve a good
qualitative description of both Mott insulator and superfluid limits and yet simple enough that it can be applied with relative ease to finite
spatially inhomogeneous bosonic optical lattice experiments.  The theory is 
a generalization of the mean-field theory of the MI-SF
phase transition in which the potential which induces coherence between different number states on a given lattice site is elevated from a
variational parameter to a fluctuating quantum variable.  
We illustrate the potential of this simple theory by applying it to uniform optical
lattice systems with constant and fully frustrated inter-site hopping parameters.

\begin{figure}
\includegraphics[clip,viewport=0 0 565 275,width=4.5in,angle=0]{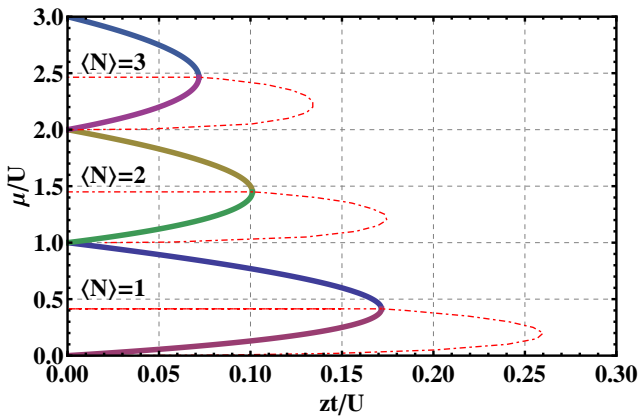}
\caption{Mean field phase diagram of the BHH. The solid lobes correspond to the MI phases, characterized by integer occupation of atomic sites.  $z$ is the coordination number, the number of neighbors of any given site. 
For sufficiently large values of $zt/U$, the system enters the SF phase. The dashed curves are contours along 
which the coherence field Berry curvature vanishes (see text).} \label{unfrphdiag}
\end{figure}

The systems in which we are interested
provide an approximate realization of the Bose Hubbard Hamiltonian (BHH) \cite{PhysRevB.40.546},
\begin{equation}
H_{BHH} =  \frac{1}{2}\sum_i Un_i(n_i-1)-\sum_i(\mu-\epsilon_i)n_i-
\sum_{\langle i,j\rangle} t_{ij}a^{\dagger}_{i}a^{\vphantom{\dagger}}_{j}.\label{BHH}
\end{equation}
The BHH provides an accurate description of cold atom systems in which the 
optical lattice potential is strong enough that only the lowest Bloch band is
significantly occupied \cite{PhysRevLett.81.3108}. In Eq. \eqref{BHH}, $a^{\dagger}_{i}$ is the boson creation operator at site $i$, $n_i=a^{\dagger}_{i}a^{\vphantom{\dagger}}_{i}$ is the number operator, $t_{ij}$ is the hopping amplitude between sites $i$ and $j$,
$U$ is the on-site interaction energy, $\mu$ is the chemical potential, and $\epsilon_i$ is an energy offset due to the trap, or to other intended or unintended local potentials.  For a translationally invariant system with nearest-neighbor hopping, mean field theory produces a phase diagram in $\mu/U$-$t/U$
space in which SF states are interrupted at small $t/U$ by a series of MI lobes centered on half-odd integer values of $\mu/U$, each
characterized by a different fixed integer value $N$ of the number of atoms per site (Fig. \ref{unfrphdiag}). 
Since $t$ decreases and $U$ increases with optical lattice potential strength, $t/U$ can be experimentally varied over a wide range.

Most studies of the BHH have focused on some regime of the model's parameter space.
For instance, in the small $t/U$ limit atom number fluctuations on a given site due to hopping can be treated as weak
perturbations. Even near the SF-MI transition strong interactions still suppress number fluctuations significantly, reducing the physically
relevant Fock subspace to two or three number states and justifying
mappings which transform the BHH into spin models that can be
attacked using a large arsenal of extensively developed techniques  \cite{Auerbach}. On the other hand, for large values of $t/U$ the interactions between cold atoms are weak enough to justify Bogoliubov's weakly interacting boson theory. 
For large numbers of atoms per unit cell one can often employ the {\em rotor approximation}, $a^{\vphantom{\dagger}}_i \cong \sqrt{\bar n} e^{i\theta_i}$ that is valid when the mean occupation number $\bar n$
at each site is so large that its relative fluctuations are small. 
The resulting Hamiltonian is a quantum phase model in which the degrees of freedom
are the phases of the superfluid at different sites. 
In this limit interactions induce phase fluctuations around a mean-field state in which
all sites adopt a common phase \c
ite{Sachdev}.  All of these approaches have disadvantages in describing realistic optical lattice experimental systems which
may have local superfluidity in one part of the system and local insulating behavior in another, 
and which typically have a mean boson number
on each lattice site of order 1 \cite{PhysRevLett.97.060403, GretchenK}. 
Our approach has goals that are similar to those of 
other complementary approximate theories \cite{PhysRevB.44.10328, PhysRevB.44.10328, PhysRevB.45.3137, PhysRevB.53.2691, PhysRevB.59.12184, PhysRevA.63.053601, PhysRevA.68.043623, PhysRevA.70.053609, Garcia-Ripoll:04, PhysRevA.71.033629, PhysRevA.73.033621, PhysRevA.74.043609, PhysRevA.75.053616, PhysRevA.79.013614, PhysRevA.79.043601, PhysRevB.76.094503, PhysRevB.77.235106, PhysRevLett.100.050404}.
We seek an approach that can adequately describe the physics of the BHH 
in both insulating and superfluid regimes and is simple enough that it can be applied in the presence of inhomogeneities.  One advantage of our approach is that we treat interactions exactly, a feature most useful in describing strongly correlated systems.

Our paper is organized as follows.  In Sec. II, we describe the flexible formalism that is the subject of this paper.
In practical applications it leads to a quadratic action for elementary excitations of uniform or non-uniform interacting bosons.
In Section III, we report on illustrative applications first to the case of a uniform BHH with constant intersite tunneling amplitudes and then,
as an example of a non-uniform system, to the case of a uniform
BHH with fully frustrated intersite tunneling amplitudes.
In Sec. IV we discuss the limitations of our theory before concluding with a brief summary.

\section{Formalism}

Our approach is based on single-site interacting boson wavefunctions $|\psi(\Sigma)\rangle$ which depend on a complex parameter $\Sigma$ and are
defined as Fock-space normalized ground states of the following single-site Hamiltonian,
\begin{equation}
h(\Sigma) = \frac{U}{2} n (n-1) - \mu n - \Sigma a^{\dagger} - \bar{\Sigma} a. \label{smallh}
\end{equation}
Notice that the potential $\Sigma$ induces coherence between single site states with different boson occupation numbers.  The 
mean field theory of the BHH SF-MI phase transition \cite{Sheshadri} can be derived by considering variational wavefunctions of the following form,
\begin{equation}
|\Psi(\Sigma)\rangle_{MF} = \prod_{i} |\psi_i\rangle.
\end{equation}
These mean-field wavefunctions do not allow for correlated inter-site fluctuations.
The mean-field ground state is determined by minimizing
\begin{equation}
E(\Sigma)\equiv \frac{\langle \Psi(\Sigma) | H_{BHH} | \Psi(\Sigma) \rangle}{\langle \Psi(\Sigma) |  \Psi(\Sigma) \rangle}
\end{equation}
with respect to the variational parameter $\Sigma$.  In the SF state $\Sigma_{MF} \ne 0$.  

Our approach is to elevate $\Sigma$ from a
variational parameter to a quantum variable with correlated spatial fluctuations by allowing it to depend on site and on imaginary time ($\Sigma
\to \Sigma_i(\tau)$) and then to construct an action $S$ which depends on these fluctuations \cite{NegeleOrland,JackiwKerman},
\begin{equation}
S=\int_{0}^{\beta}d\tau \left[ \; \sum_{i} {}_i\langle \psi(\Sigma_i(\tau))|\partial_\tau\psi(\Sigma_i(\tau)) \rangle_i + E[\Sigma]\; \right].
\label{action}
\end{equation}
Here at each instant of imaginary time
\begin{equation}
E[\Sigma]= \frac{\langle\Psi[\Sigma]|H_{BHH}|\Psi[\Sigma]\rangle}{\langle \Psi[\Sigma] |  \Psi[\Sigma] \rangle}, \label{enfnc}
\end{equation}
and the correlated product state is given by
\begin{equation}
|\Psi[\Sigma]\rangle = \prod_{i} |\psi(\Sigma_i)\rangle_{i}.
\end{equation}
In practice the action can be evaluated analytically only if the coherence fields are expanded to leading order 
around their mean-field values.  
The main advantage of this approach, as stressed above, is its convenience in practical calculations, especially for non-uniform systems.  Before we elaborate on this point, we 
examine some formal properties of our single site interacting boson wavefunctions $|\psi(\Sigma)\rangle$.

\begin{figure*}
\centering
\mbox{ {\scalebox{0.75}[0.75]
{\includegraphics*[clip,viewport=45 15 355
250,width=5.35in,height=4.75in,angle=0]{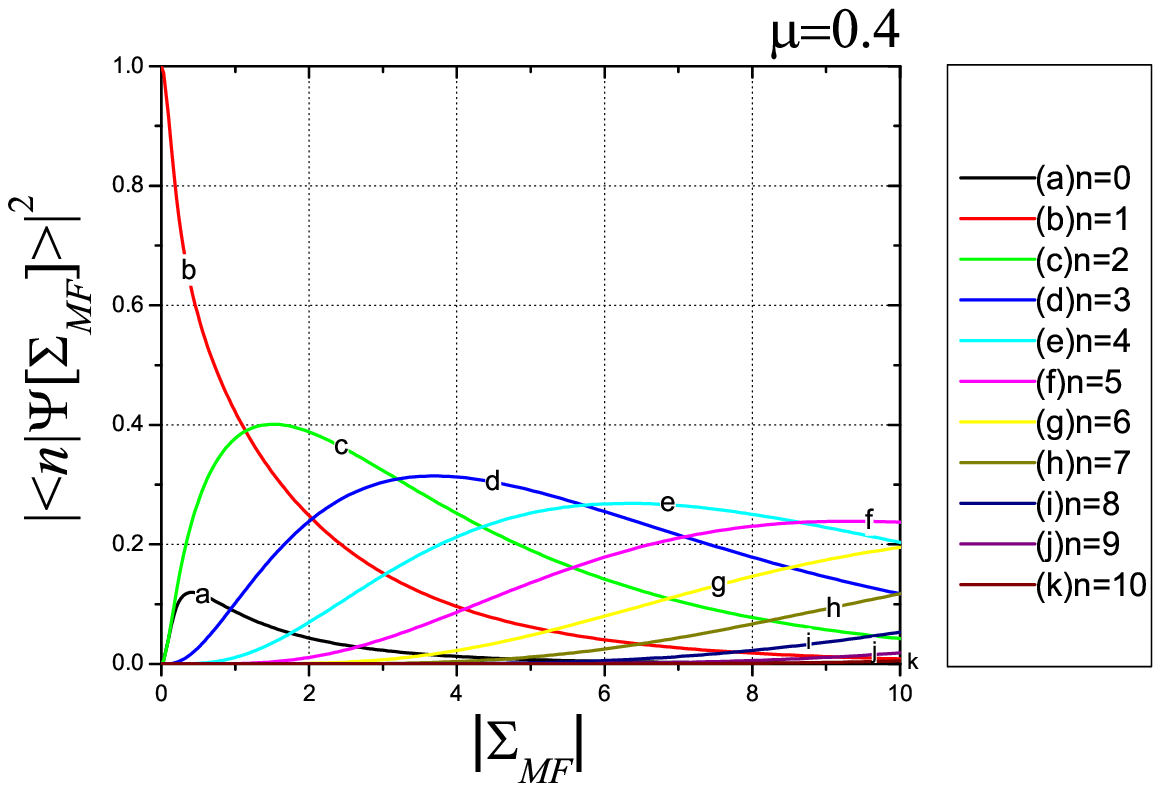}}} {\scalebox{0.75}[0.75]{\includegraphics*[clip,viewport=50 15 350 250,width=4.9in,height=4.75in,angle=0]{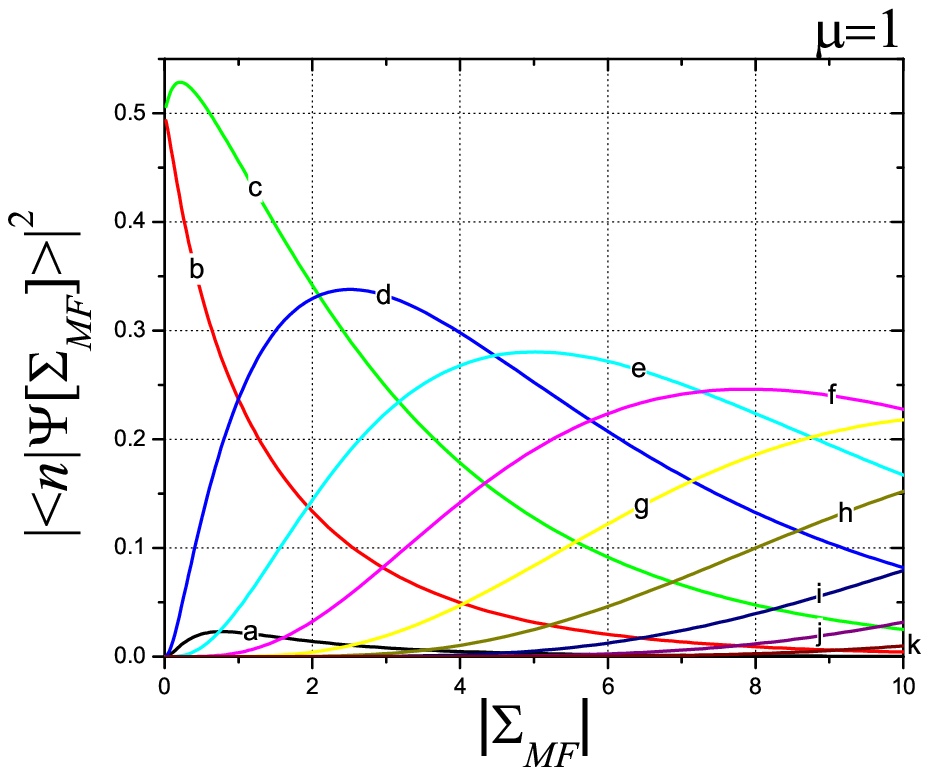}}}}
\caption{The overlap of our interacting wavefunctions with Fock number states, $|\langle n|\psi(\Sigma_{MF})\rangle|^2$.  Deep inside the SF phase (large $|\Sigma_{MF}|$), more number states come into play, the result of strong atom number fluctuations.  At $\mu=0.4U$, only one Fock state, {\em viz.} $|n=1\rangle$, is dominant close to the phase boundary (small $|\Sigma_{MF}|$), whereas there are two, $|n=1\rangle$ and $|n=2\rangle$, at $\mu=U$, in accordance with Fig. \ref{unfrphdiag}.}\label{numstates}
\end{figure*}

\subsection{Formal Properties of the Wavefunction $|\psi(\Sigma)\rangle$}

One method of characterizing the Fock space wavefunctions $|\psi(\Sigma)\rangle$ is to consider their expansion in terms of number eigenstates,
\begin{equation}
|\psi(\Sigma)\rangle = \sum_{n=0}^{\infty} c_n(|\Sigma|) \exp(i n \phi_{\Sigma}) |n\rangle, \label{ocnumexp}
\end{equation}
where we have noted that the magnitude of the expansion coefficients depends only on $|\Sigma|$,
and defined $\phi_{\Sigma}$ as the phase of
$\Sigma$. In Fig. \ref{numstates} we plot $c_n^2=|\langle n|\psi(\Sigma)\rangle|^2$ {\em vs.} $|\Sigma|$ for a variety of $n$ values for both
$\mu/U=0.4$, which falls inside the $N=1$ MI lobe and for $\mu/U=1.0$ at the boundary of the $N=1$ and $N=2$ MI lobes.  This plot illustrates why spin model
approximations to the BHH are justified close to the transition (for small values of $|\Sigma|$), since a 
small number of number states dominate.
As one goes deeper into the SF region, the potential $\Sigma$ induces coherence amongst more number states
and spin-model approximations will fail. 

The mean field state is characterized by a time independent field $\Sigma_{MF}$ that minimizes the energy,
\begin{equation}
\frac{\partial E}{\partial \Sigma_i}\Big |_{MF} = 0.\label{MFstate}
\end{equation}
(The first term of \eqref{action} does not contribute to the action if $\Sigma$ is time-independent.)
Quantum fluctuations are incorporated by employing
a Gaussian-fluctuations approximation, $\Sigma_i = \Sigma_{MF} + z_i$. 
Since the time dependence
comes from the fields only, we expand the time derivative in the Berry phase term as
\begin{equation}
\partial_\tau =
\frac{\partial}{\partial\Sigma_i}\dot{\Sigma}_i + \frac{\partial}{\partial\bar\Sigma_i}\dot{\bar\Sigma}_i = {\partial_{\Sigma_i}}\dot{z_i} +
{\partial_{\bar\Sigma_i}}\dot{\bar z}_i.\label{timederivative}
\end{equation}
Substituting this in the action enables us to rewrite the first term of \eqref{action}, the Berry phase term, as $\mathcal{C}_i(\Sigma_{MF})\bar z_i\dot{z}_i$, where the gauge invariant Berry curvature of site $i$ $C_i$, evaluated at $\Sigma_{MF}$, is given by \cite{MVBerry}
\begin{equation}
\mathcal{C}_i(\Sigma_{MF}) = \bigg\langle\frac{\partial\psi_i}{\partial\bar\Sigma_i}\bigg|
\frac{\partial\psi_i}{\partial\Sigma_i}\bigg\rangle -
\bigg\langle\frac{\partial\psi_i}{\partial\Sigma_i}\bigg|
\frac{\partial\psi_i}{\partial\bar\Sigma_i}\bigg\rangle,\label{berrydef}
\end{equation}
The Berry phase contribution to the action 
specifies the quantization condition of our fluctuating variables and plays an 
essential role in determining elementary excitation energies.  

The energy functional, Eq. \eqref{enfnc}, can also be expanded around its mean-field value,
\begin{eqnarray}
E[\Sigma]&=&E_{MF} + \frac{1}{2}  \sum_{ij} \Big[ \frac{d^2E}{d\Sigma_id\Sigma_j}\bigg|_{MF}z_iz_j\nonumber\\&+&
\frac{d^2E}{d\bar\Sigma_id\Sigma_j}\bigg|_{MF}\bar z_iz_j + c.c. \Big] ,\label{enexp}
\end{eqnarray}
where $i$ and $j$ stand for lattice sites and $h.c.$ for complex conjugate. 
Combining the Berry phase term with the second order contribution to the energy functional $E^{(2)}$, 
we construct a quadratic action from which we can use to calculate the elementary 
excitations. $S \cong S_{MF} + S^{(2)}[\bar z_i, z_i]$, where
\begin{equation}
S^{(2)}[\bar z_i, z_i]=\int_{0}^{\beta}d\tau\sum_{i} \left[\mathcal{C}_i(\Sigma_{MF})\bar z_i\dot{z}_i + E^{(2)}[\Sigma]\right].\label{action2}
\end{equation}
$E^{(2)}[\Sigma]$ contains all the second order terms of \eqref{enexp}.

\subsection{Single Site States at Large $n$.}

Further insight into the properties of the single site Hamiltonian,  Eq.~\eqref{smallh}, can be obtained by examining the quantum phase model, 
\begin{equation}
h(\Sigma) = \frac{U}{2} n (n-1) - \mu n - 2\sqrt{n}|\Sigma|\cos(\theta-\theta_\Sigma), \label{smallh2}
\end{equation}
that is derived from Eq.~\eqref{smallh} by letting $\Sigma \rightarrow |\Sigma |e^{i\theta_\Sigma}$, and employing the rotor approximation $a\rightarrow \sqrt{n}e^{i\theta}$, which is valid when number density fluctuations are small.
In the rest of this section we write $\Sigma$ for $|\Sigma|$.
Assuming that quantum fluctuations are small allows us to determine the average atom number $n_o \sim (\Sigma/U)^{2/3}$ and phase $\theta_o=\theta_\Sigma$ by minimizing $h(\Sigma)$ with respect to $n$ and $\theta$. In addition, we can expand Eq.~\eqref{smallh2} to second order about the extrema $n_o$ and $\theta_\Sigma$ to arrive at the quadratic Hamiltonian,
\begin{equation}
h(\Sigma) = const + \frac{3U}{4} (n-n_o)^2 + \frac{\Sigma^{4/3}}{U^{1/3}}(\theta-\theta_\Sigma)^2. \label{quadham}
\end{equation}
Phase and atom number are conjugate variables and we therefore 
recognize \eqref{quadham} as a quantum
harmonic oscillator Hamiltonian.  Using this analogy we find 
the energy level spacing $\omega=\sqrt{3}U^{1/3}\Sigma^{2/3}$, mass $m=2/3U$,
typical density fluctuation $\delta n \equiv \langle (n-n_o)^2 \rangle^{1/2} 
\sim (\Sigma/U)^{1/3}$ and typical phase fluctuation
$\delta \theta \equiv \langle (\theta-\theta_\Sigma)^2 \rangle^{1/2} \sim (U/\Sigma)^{1/3}$.
We see that density fluctuations are suppressed and phase fluctuations enhanced as 
$\Sigma_{MF}$ approaches $0$ at the MI transition boundary.

\begin{figure}
\includegraphics[clip,viewport=15 0
275 225,width=3.5in,angle=0]{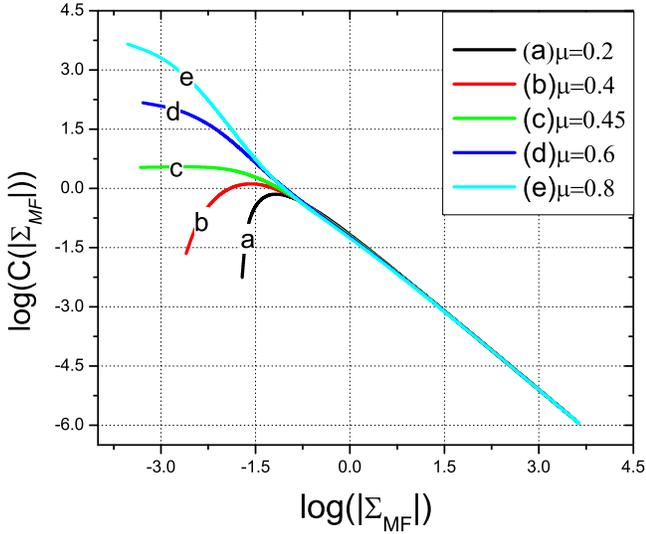}
\caption{A log-log plot of the Berry curvature as a function of the coherence field $|\Sigma_{MF}|$. Note the power 
law decay rule $\mathcal{C}_{i}(\Sigma_{MF}) \backsim |\Sigma_{MF}|^{-4/3}$ 
at large $|\Sigma_{MF}|$ values predicted by Eq. \eqref{4third}. This plot is calculated from Eq. \eqref{berrydef}. The chemical potentials $\mu$ are measured in units of $U$.} \label{berryplot}
\end{figure}

The harmonic oscillator analog, based on the identifications $p\leftrightarrow\delta n $ and $q\leftrightarrow\delta\theta$, can also be used to derive an expression for the on-site 
Berry curvature that is valid at large $\Sigma$.  The Berry curvature is determined by the 
dependence of the single-site wavefunction on the magnitude and phase of $\Sigma$.
We therefore consider the influence of perturbations on the eigenstates of 
Eq.~\eqref{quadham},
\begin{equation}
h' = \frac{p^2}{2m} + \frac{1}{2}m\omega^2q^2 - \lambda_pp-\lambda_qq= h-\lambda_pp-\lambda_qq. \label{hwithpert}
\end{equation}
To first order in $\lambda_p$ and $\lambda_q$ the ground state of $h'$ is given by
\begin{equation}
|\phi_o\rangle ' = |\phi_o\rangle + \sum_{n \neq o} \frac{\langle
\phi_n|\lambda_pp+\lambda_qq|\phi_o\rangle}
{\varepsilon_n-\varepsilon_o}|\phi_n\rangle.\label{phiprime}
\end{equation}
where $\varepsilon_n$ and $|\phi_n\rangle$ are the eigenvalues and eigenfunctions of 
$h$. 
It follows that the harmonic oscillator 
Berry curvature is given by (cf. Eq.~\eqref{berrydef})
\begin{eqnarray}
\mathcal{C} &=& \text{Im}\left[\left\langle\frac{\partial\phi}{\partial\lambda_2}\bigg|
\frac{\partial\phi}{\partial\lambda_1}\right\rangle-
\left\langle\frac{\partial\phi}{\partial\lambda_1}\bigg|
\frac{\partial\phi}{\partial\lambda_2}\right\rangle\right]\label{berry}\\
 &=&\sum_{n\neq o}\text{Im}\left[\frac{\langle \phi_o|p|\phi_n\rangle\langle\phi_n|q|\phi_o\rangle-\langle
\phi_o|q|\phi_n\rangle\langle\phi_n|p|\phi_o\rangle}
{\left(\varepsilon_n-\varepsilon_o\right)^2}\right].\nonumber
\end{eqnarray}
The matrix elements in this expression can be evaluated using $[h,p]=im\omega^2q$ and $[h,q]=-ip/m$ to find 
\begin{equation}
\mathcal{C}(\Sigma)=\frac{1}{\omega^2},\label{4third}
\end{equation}
and hence for our single-site states $C \sim \Sigma^{-4/3}$ at large $\Sigma$.
This result is confirmed by the plot (Fig. \ref{berryplot}) of Berry curvature 
values numerically obtained from Eq. \eqref{berrydef}.
 
\section{Application to the Bose Hubbard Hamiltonian}

\begin{figure*}
\begin{center}
\mbox{ \subfigure[ Dispersion inside the MI region.] {\scalebox{0.75}[0.75]
{\includegraphics*[clip,viewport=25 0
550 400,width=3.25in,height=3.5in,angle=0]{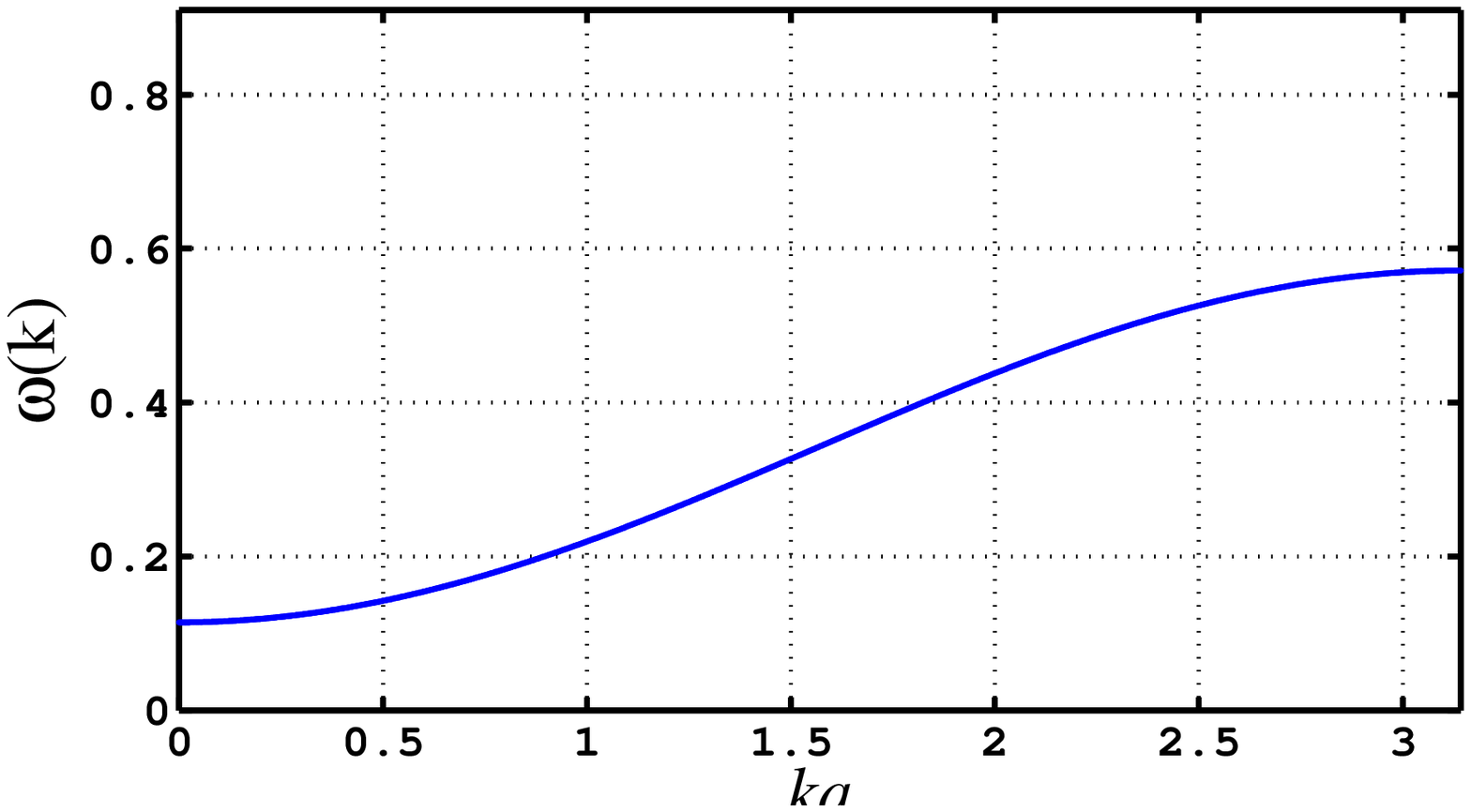}}}
\subfigure[ Dispersion at the phase boundary.] {\scalebox{0.75}[0.75]{\includegraphics*[clip,viewport=75
0 545 400,width=3.0in,height=3.25in,angle=0]{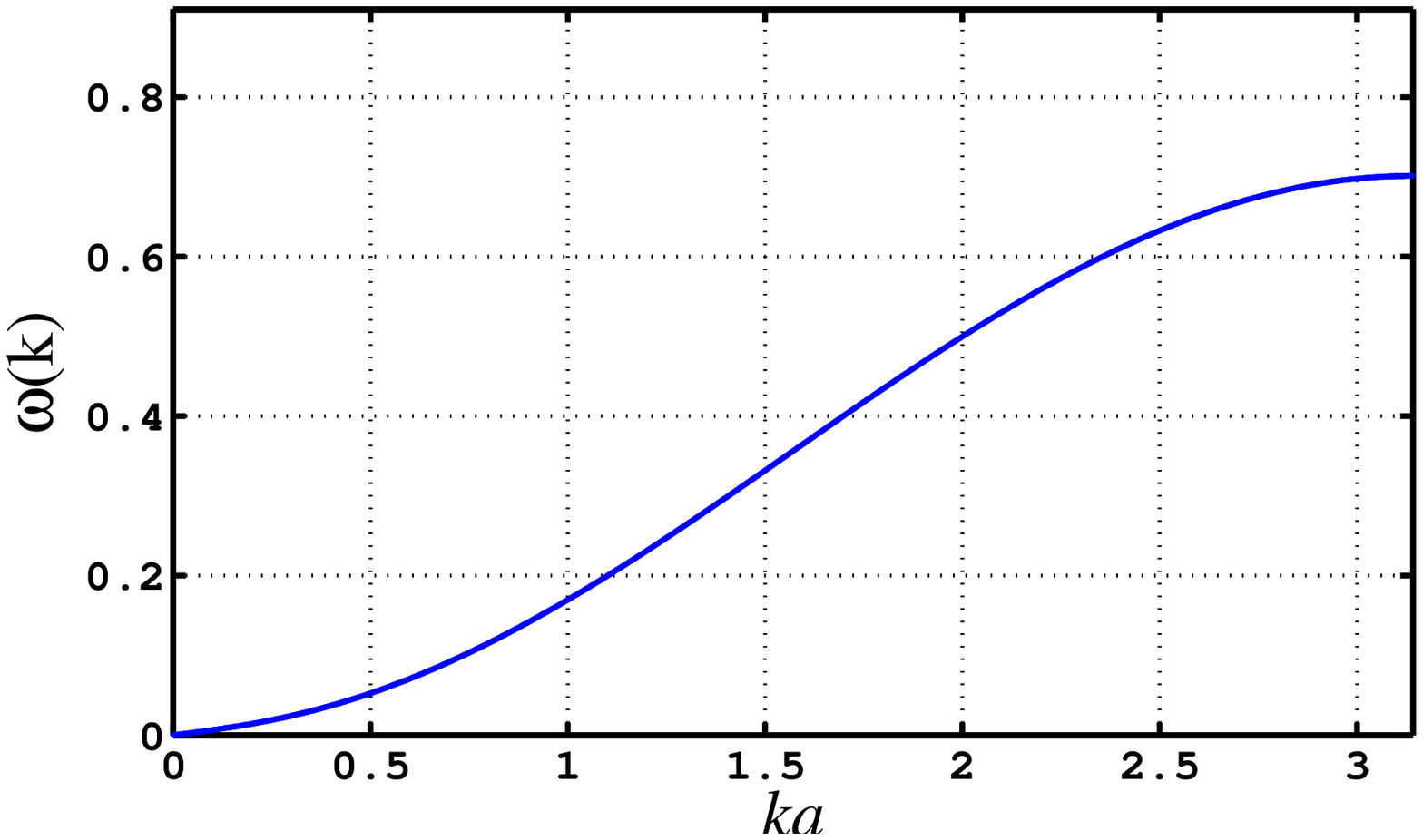}}}
\subfigure[ Dispersion inside the SF phase.] {\scalebox{0.75}[0.75]{\includegraphics*[clip,viewport=75 0
545 400,width=3.0in,height=3.25in,angle=0]{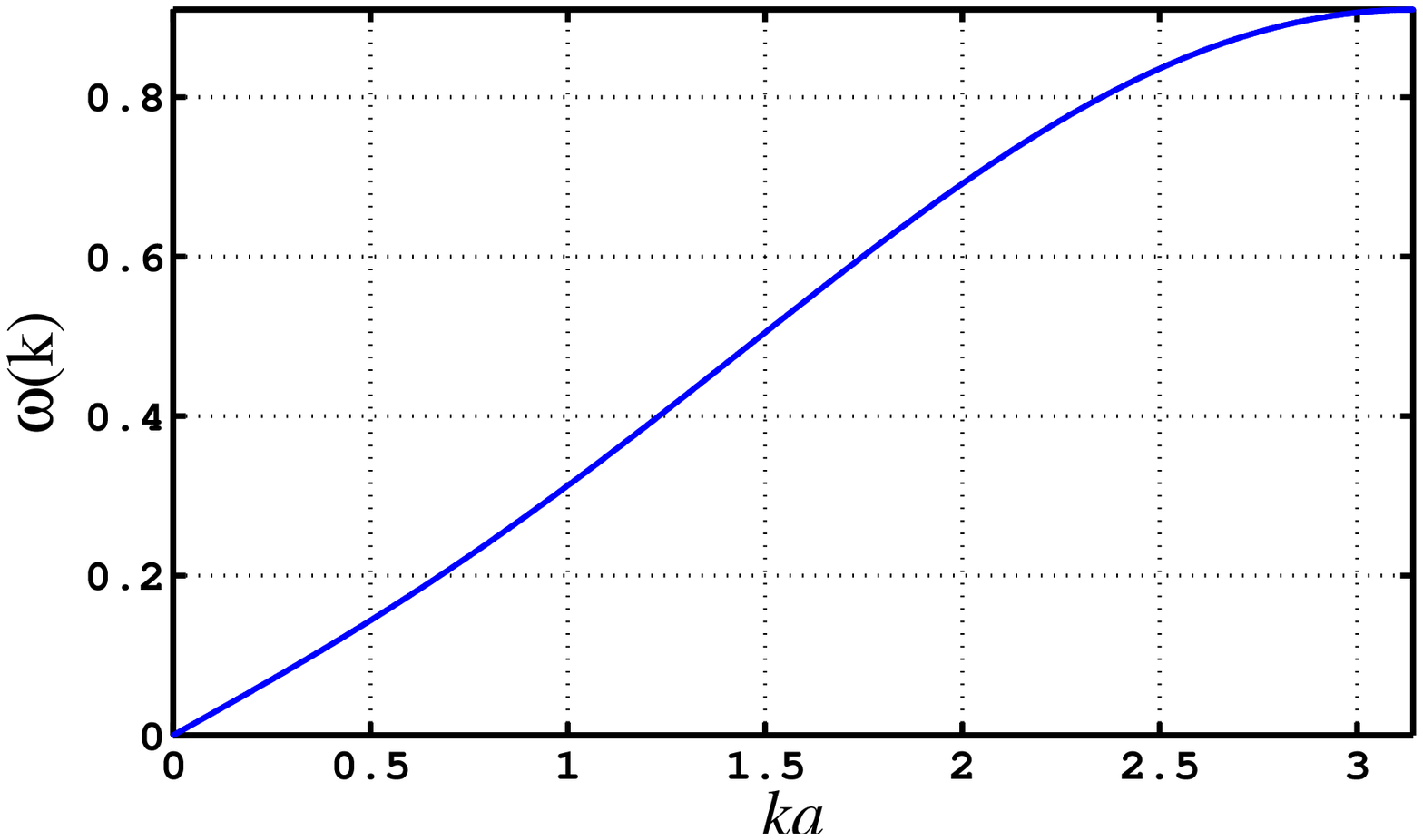}}}}
\caption{Elementary excitation energy $\omega(k)$ vs $k$ for a two dimensional 
uniform BHH along the line, $k \equiv k_x=k_y$. Note that the spectrum is gapped in
the insulating phase, and gapless at and beyond the phase boundary. Also note that for small $\mathbf{k}$, the dispersion is linear in the SF region, in accordance with Goldstone's theorem.  In this approximate theory the dispersion is quadratic at the phase boundary.} \label{dispfigs}
\end{center}
\end{figure*}

We now test the theory's practical utility, first by applying it to the BHH for a uniform optical lattice, and secondly by applying it to the fully frustrated BHH which has four atoms per unit cell and therefore introduces inhomogeneity. Separating $\Sigma_i$ into its mean-field and fluctuation contributions, the single site Hamiltonian $h_i$ becomes
\begin{equation} 
h_i=\frac{n_i(n_i-1)}{2}-\mu n_i-a^{\dagger}_{i}\Sigma_{MF}
-a^{\vphantom{\dagger}}_{i}\bar\Sigma_{MF} -a^{\dagger}_{i}z_i -a^{\vphantom{\dagger}}_{i}\bar z_i.\label{smallh3}
\end{equation}
The fluctuations are treated perturbatively. We write 
\begin{equation}
|\psi\rangle_i = |\psi_o\rangle_i + |\psi_\Sigma
\rangle_iz_i + |\psi_{\bar\Sigma}\rangle_i
\bar z_i,\label{pertst}
\end{equation}
where 
\begin{equation}
|\psi_\Sigma \rangle_i= \frac{\partial|\psi\rangle_i}{\partial\Sigma_i} = -\sum_{n \neq 0} \frac{{}_i\langle \psi_n|a^{\dagger}_{i}
|\psi_o\rangle_i}{E_o-E_n}|\psi_n\rangle_i. \label{psiderv}
\end{equation}
Here $|\psi_n\rangle_i$ and $E_n$ are the eigenstates and the energy levels of the unperturbed on-site Hamiltonian. With Eq. \eqref{psiderv}, the Berry curvature can be calculated from Eq. \eqref{berrydef} quite easily. The energy of the system, given by Eq. \eqref{enfnc}, can also be expanded (to quadratic order) in terms of the fluctuating fields (cf. Eq. \eqref{enexp}),
\begin{equation}
E = E^{(o)} + \sum_iE_{\Sigma_i}z_i + \sum_{ij}\left(E_{\Sigma_i\Sigma_j}z_iz_j +
E_{\bar\Sigma_i\Sigma_j}\bar z_iz_j\right) + h\cdot
c \cdot,\label{toten}\end{equation}
where we naturally identify 
\begin{equation}
E_{\Sigma_i} = \frac{\partial E}{\partial\Sigma_i}\bigg|_{MF}
\textrm{, } E_{\Sigma_i\Sigma_j} = \frac{1}{2}\frac{\partial^2E}{\partial\Sigma_i\partial\Sigma_j}\bigg|_{MF} \textrm{,}\cdot\cdot\cdot.\label{scndev}
\end{equation} 
The indices $i$ and $j$ refer to either same site or neighboring sites. For a uniform lattice, both the Berry curvature and the energy derivatives are independent of site indices. The mean field state is determined by setting the first derivative terms to 
zero.  The MI phase boundary is defined by the largest value of $t/U$ for a given $\mu/U$ 
for which the energy is miminized by $|\Sigma| = 0$ on all sites.
For the uniform BHH this procedure reproduces the familiar phase diagram plotted in Fig. \ref{unfrphdiag}.

\subsection{Elementary Excitations in a Uniform Lattice}

To determine the elementary excitations of the BHH \cite{PhysRevA.63.053601}, 
we turn to the second order action, Eq. \eqref{action2}, where we use Eq. \eqref{toten} for the energy functional. For a uniform lattice, we Fourier-transform the fluctuations $z_i$,
\begin{equation}
z_i=\frac{1}{\beta\sqrt{N}}\sum_{k,n}z_{kn}e^{i(\mathbf{k\cdot
r}_i-\omega_n\tau)},\label{ft}
\end{equation}
where $\beta$ is the inverse temperature, $\omega_n$ are the Matsubara frequencies and $N$ here is the number of sites in the lattice. The resulting action is 
\begin{eqnarray}
S^{(2)}[\bar zz]&=&\frac{1}{\beta}\sum_{k,n}[-i\omega_n \, \mathcal{C}(\Sigma_{MF})\bar z_{kn}z_{kn}\nonumber+ \bar z_{kn} A_k z_{kn}
\\&+&z_{kn} \bar B_k z_{-k-n} +
\bar z_{kn}B_{k} \bar z_{-k-n}]\nonumber\\ \label{themess} \\&=&\frac{1}{\beta}\sum_{k,n \geq 0}
\begin{array}{cc}[\bar z_{kn} & z_{-k-n}]\\
\vphantom{a}\end{array} \begin{array}{cc} M(\mathbf{k},n) \end{array} \left[\begin{array}{cc}z_{kn} \\
\bar z_{-k-n}
\end{array} \right]\nonumber
\end{eqnarray}
where,
\begin{align}
&M(\mathbf{k},n) = \left[\begin{array}{cc}-i\omega_n\mathcal{C}(\Sigma_{MF})+A_{k} &\bar 
B_{k}\\B_{k} & i\omega_n\mathcal{C}(\Sigma_{MF})+A_{k}\end{array} \right],\nonumber\\
&A_k=
\frac{1}{2}\frac{\partial^2E}{\partial\bar\Sigma_i\partial\Sigma_i}\bigg|_{MF}
+\frac{\partial^2E}{\partial\bar\Sigma_i\partial\Sigma_j}\bigg|_{MF}(\cos(k_xa)+\cos(k_ya)),\nonumber\\
&B_k=
\frac{\partial^2E}{\partial\Sigma_i\partial\Sigma_i}\bigg|_{MF}
+\frac{\partial^2E}{\partial\Sigma_i\partial\Sigma_j}\bigg|_{MF}(\cos(k_xa)+\cos(k_ya)),\label{MAandB}
\end{align}
and $a$ is the lattice constant. Here, the indices $i$ and $j$ denote neighboring sites. By setting the determinant of the matrix $M(\mathbf{k},n)$ in Eq.~\eqref{themess} to zero, we
obtain an expression for the excitation spectrum,
\begin{equation}
\omega(k) =
\frac{ \sqrt{A_k^2-|B_k|^2}}{\mathcal{C}(\Sigma_{MF})}.\label{dispeqn}
\end{equation}
In accordance with Goldstone's theorem, Eq. \ref{dispeqn} yields
a gapless Goldstone mode in the SF phase with linear dispersion at long 
wavelengths (Fig. \ref{dispfigs}c).
As the MI phase boundary is approached and crossed, excitations become 
more localized and the mode dispersion weakens (Fig. \ref{dispfigs}b) and (Fig. \ref{dispfigs}a). 
At certain points in the phase diagram the Berry curvature $\mathcal{C}(\Sigma_{MF})$ vanishes (Fig. \ref{unfrphdiag})
and our theory of the excitation spectrum becomes unreliable.
We return to this point in the discussion section.

\begin{figure}
\begin{picture}(250,240)
\put(30,10){\line(0,1){220}} \linethickness{1mm}
\put(90,10){\line(0,1){220}} \thinlines
\put(150,10){\line(0,1){220}} \linethickness{1mm}
\put(210,10){\line(0,1){220}} \thinlines
\put(10,30){\line(1,0){220}} \put(10,90){\line(1,0){220}}
\put(10,150){\line(1,0){220}} \put(10,210){\line(1,0){220}}
\put(70,70){\dashbox{2}(100,100)} \put(80,152){$A$}
\put(140,152){$B$} \put(140,92){$C$} \put(80,92){$D$}
\put(200,152){$A$} \put(20,152){$B$} \put(20,92){$C$}
\put(200,92){$D$} \put(140,212){$C$} \put(80,212){$D$}
\put(80,32){$A$} \put(140,32){$B$} \put(74,120){$-t$}
\put(155,120){$t$} \put(118,155){$t$} \put(118,78){$t$}
\end{picture} \caption{Fully Frustrated Lattice:
The thick bonds are the links with hopping amplitude
$t_{ij}=-t$ that frustrate inter-site coherence. 
The dashed box contains a unit cell of
the frustrated lattice.} \label{Latt}
\end{figure}
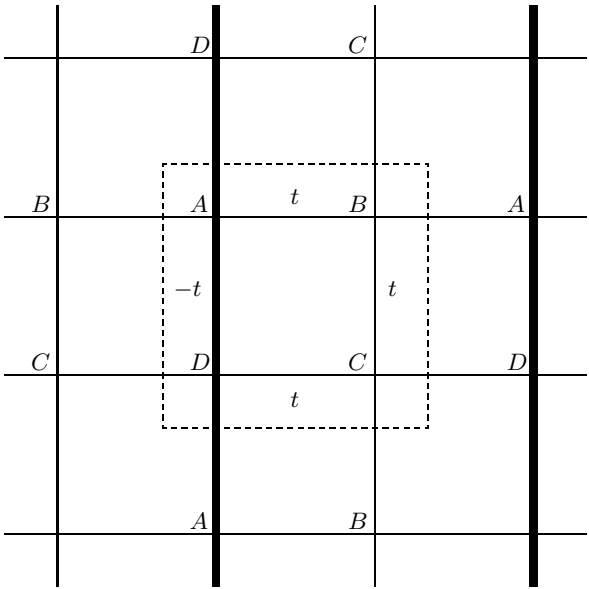
\subsection{Fully Frustrated Lattices}

\begin{figure*}
\begin{center}
\mbox{ \subfigure[Excitation modes inside the MI phase. Each mode is doubly degenerate.]{\scalebox{1.25}[1.25]{\includegraphics*[clip,viewport=0 0 600
350,width=2.750in,height=2.0in,angle=0]{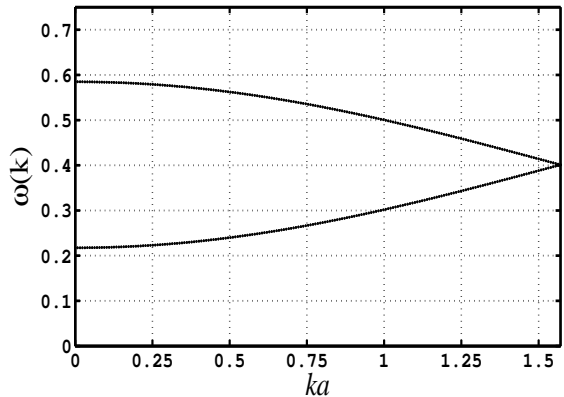}}} \subfigure[Excitation modes inside the SF phase.]{\scalebox{1.25}[1.25]{\includegraphics*[clip,viewport=0 0 600 350,width=2.750in,height=2.0in,angle=0]{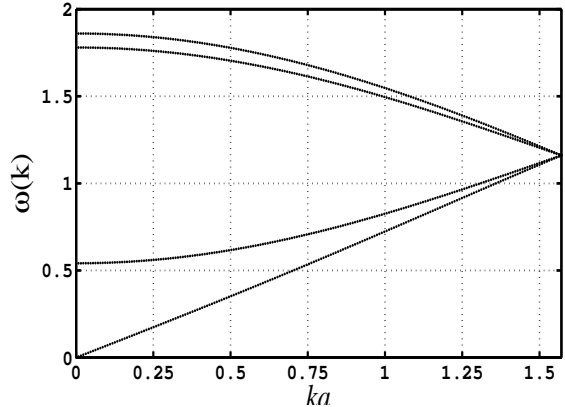}}}} 
\caption{Elementary excitations of the fully frustrated lattice plotted along $k\equiv k_x=k_y$. The lowest dispersion is gapped inside the insulator, and gapless beyond the phase
boundary. Again, obeying Goldstone's theorem, the lowest dispersion in the SF phase is gapless, and linear at small $k$.} \label{frustdisp}
\end{center}
\end{figure*}

As mentioned above, the theory outlined in Sec. II is designed with inhomogeneous systems in mind, and is general enough to be applied to systems where the translational symmetry is reduced, or altogether lost. We demonstrate this with the case of a two-dimensional uniform lattice in which the tunneling amplitude sign alternates in one direction (See Fig. \ref{Latt}). This hopping model corresponds to a half-flux quantum per square lattice 
plaquette and is referred to as a \emph{fully frustrated} BHH \cite{PhysRevLett.95.010401}. For this case we allow translational symmetry to be broken by doing the mean-field minimization for a lattice with four sites per unit cell.  
The quadratic theory is similarly modified, with the $M$ matrix in Eq. \eqref{MAandB} enlarging to an $8$ by $8$ matrix. In general, for the case of an entirely inhomogenous lattice with $N$ sites, the mean field state is determined by $N$ linearly coupled equations, while the elementary excitations are obtained by performing a Bogoliubov diagonalization on the $2N$ by $2N$ matrix resulting from the second order action \eqref{action2} while preserving the bosonic commutation relations \cite{zulicke, BlaizotRipka} (This procedure is discussed in the Appendix.) 

A detailed mean field study of the quantum phase transitions in the fully frustrated lattice has been carried out elsewhere \cite{frustMF}. Here we
apply Eq. \eqref{MFstate} on each site of a plaquette to find the mean field values of the fields, $\Sigma_A, \Sigma_B, \Sigma_C \text{ and }
\Sigma_D$. The elementary excitations of the system follow from the second order action, Eq. \eqref{action2}. We take advantage of the (reduced) translational symmetry by Fourier decomposing the fluctuation fields at each site,
\begin{equation}
z_{sl}(\mathbf{r},\tau)=\frac{1}{\beta\sqrt{N}}\sum_{k}z_{sk}(\tau)e^{i\mathbf{k\cdot
r}_l},\label{ffz}
\end{equation}
where $l$ refers to plaquettes and $s$ to
the four sites ($A$, $B$, $C$ and $D$) within a given plaquette. The Berry
phase term, calculated at the mean field state defined by the above four fields, becomes 
\begin{equation}
\sum_{sk}\mathcal{C}_s\bar z_{sk}(\tau)\partial_\tau z_{sk}(\tau),\label{ffberry}
\end{equation}
while the energy term contains all the second order deviations of the energy about this mean field state. Replacing $z_{sk}(\tau) \rightarrow \sqrt{\mathcal{C}_s}z_{sk}(\tau)$ reduces the action
into the standard form
\begin{equation}
S^{(2)}=\int_{0}^{\beta}d\tau \sum_{k,s} \left(\bar
z_{sk}(\tau)\partial_\tau z_{sk}(\tau) + E^{(2)}[\bar z_{sk}(\tau),z_{sk}(\tau)]\right).
\label{standard}
\end{equation}
Having elevated the fluctuations into quantum variables (obeying bosonic commutation relations), we are now in a position to employ the transformation $\vec z = \mathcal{B} \vec v$, where $ \vec z = \left[\begin{array}{c} z_{sk}(\tau) \\ \bar z_{s-k}(\tau) \end{array} \right],$ that preserves the commutation relations, and perform Bogoluibov diagonalization \cite{zulicke, BlaizotRipka} to transform
the action into
\begin{equation}
S^{(2)}=\int_{0}^{\beta}d\tau \sum_k \left(\bar v_k\partial_\tau
v_k + \omega(k)\bar v_k v_k \right),\label{frAct}
\end{equation}
from which we can easily identify $\omega(k)$ as the excitation spectrum. A few sample plots of the dispersion are shown in Fig. \ref{frustdisp}.

\section{Discussion}

As demonstrated above, the Berry phase is the critical ingredient in formulating the quantum theory and calculating the elementary excitations from it. As such, the excitation spectra one obtains become unreliable if and when the Berry curvature vanishes. The dashed curves in Fig. \ref{unfrphdiag} trace the contours in the phase diagram where the Berry curvature, Eq. \eqref{berrydef}, of the uniform BHH becomes zero. One point of view towards restoring the quantum theory is to retain terms that are second order in time derivative in deriving the action, Eq. \eqref{action} \cite{PhysRevD.19.2349,Shankar}. Here, we content ourselves to exploring the consequences of a vanishing Berry curvature to our theory. To have a better understanding of these special points, we focus on the MI phases, where the Berry curvature is explicitly given by 
\begin{equation}
\mathcal{C}=\frac{n+1}{\left(\frac{\mu}{U}-n\right)^2}-
\frac{n}{\left(n-1-\frac{\mu}{U}\right)^2}.
\end{equation}
For $n=1$, for instance, $\mathcal{C}$ vanishes at $\mu/U=\sqrt{2}-1$, corresponding to a set of points in the phase diagram with particle-hole symmetry. In other words, in our formalism, there is no distinction between the definitions for the annihilation and creation operators that emerge from quantizing the fluctuations, rendering their commutator zero. This is consistent with the results of Ref. \cite{PhysRevB.40.546}, where quantum phase transitions at the tips of the MI phases, the multicritical points with particle-hole symmetry, belong to the universality class of the $\mathrm{XY}$ model, where the time derivative is second order, in contrast to anywhere else across the phase boundary where it is first order, and there is an absence of particle-hole symmetry \cite{PhysRevB.40.546,Sachdev}. 

In summary, the relative ease with which the above theory has determined the basic properties of the fully frustrated optical lattice higlights its main focus: inhomogenous systems, such as optical lattices in symmetry breaking harmonic traps, or experimental set-ups with controlled disorder, which have generated lots of interest lately \cite{disorder}. The advantage of our formalism is its broad applicability and accuracy despite its simplicity, which lends to its convenience. It may also be interesting to modify the theory developed here for other lattice Hamiltonians, such as those involving next neighbor interactions, where interesting phases such as charge density waves and supersolids are predicted \cite{PhysRevB.52.16176}. 

\begin{acknowledgments}
This work is supported by Welch Foundation.
\end{acknowledgments}

\appendix
\section*{Appendix}

In this section, we discuss some of the diagonalization procedure we employed above to arrive at Eq. \eqref{frAct}. The need to perform Bogoliubov diagonalization of many degrees of freedom arises in most considerations of many body problems, especially inhomogenous ones, and there are now a number of sources in the literature one can consult for more details \cite{zulicke,BlaizotRipka}. Here we give a more succint summary. Let $$ \vec z = \left[\begin{array}{c} z_{i} \\ \bar z_{i} \end{array} \right] = \left[\begin{array}{c} z_{1} \\ z_{2} \\\vdots\\z_{N} \\ \bar z_{1} \\ \bar z_{2}\\ \vdots\\\bar z_{N} \end{array} \right],$$
where $N$ is the total number of sites in the optical lattice. As mentioned above, these fluctuations are now quantum variables, and thus obey bosonic commutation relations, $\left[\vec z, \vec z^\dagger\right]=\mathcal{I}$, where $\mathcal{I} = \left[\begin{array}{cc}\mathbbm{1} &\mathrm{0}\\
\mathrm{0} & -\mathbbm{1}\end{array} \right],$ with $\mathbbm{1}$ and $\mathrm{0}$ being the $\mathrm{N}\times \mathrm{N}$ identity and zero matrices, respectively.

We introduce the Hamiltonian $\mathcal{H}$ such that second order energy terms of the action, such as the one in Eq.\eqref{standard}, can be written as $E^{(2)}[\bar z_{i},z_{i}] = \vec z^{\dagger} \mathcal{H}\vec z.$ Our goal is to perform a canonical transformation $\vec z = \mathcal{B} \vec v$  that diagonalizes the Hamiltonian $\mathcal{H}$ while preserving the bosonic commutation relations, 
\begin{eqnarray}
\mathcal{B}^\dagger\mathcal{H}\mathcal{B}=\mathcal{D}\nonumber\\
\left[\vec v, \vec v^\dagger\right]=\mathcal{I},\label{cantran}
\end{eqnarray}
where $\mathcal{D}=\{\epsilon_1,\epsilon_2,\cdot\cdot\cdot,\epsilon_{2N}\}$ are the eigenmodes of interest. Using Eq. \eqref{cantran} alongside the identities $\mathcal{B}^\dagger\mathcal{I}\mathcal{B}=\mathcal{I}$ (which follows from the commutation relations) and $\mathcal{I}=\mathcal{I}^{-1}$, we obtain the expression 
\begin{equation}
\tilde{\mathcal{H}}\mathcal{B} = \tilde{\mathcal{D}}\mathcal{B},\label{eigeqn}
\end{equation} where $$\tilde{\mathcal{D}}\equiv\mathcal{I}^{-1}\mathcal{D}=\{\epsilon_1,\epsilon_2,\dots,\epsilon_N,
-\epsilon_{N+1},-\epsilon_{N+2},\dots,-\epsilon_{2N}\}.$$ If we consider each vector $\vec b_i$ comprising the matrix $\mathcal{B}$, we see that Eq.\eqref{eigeqn} is essentially an eigenvalue problem, where the eigenvalues are particle-hole pairs ($\epsilon_i=-\epsilon_{i+N}$), and $\epsilon_i$, where $1\leq i\leq N$ are the fundamental excitation modes of the theory such as those plotted in Fig. \ref{frustdisp}.

\bibliographystyle{apsrev}
\bibliography{BHreferences}

\begin{thebibliography}{39}
\expandafter\ifx\csname natexlab\endcsname\relax\def\natexlab#1{#1}\fi
\expandafter\ifx\csname bibnamefont\endcsname\relax
  \def\bibnamefont#1{#1}\fi
\expandafter\ifx\csname bibfnamefont\endcsname\relax
  \def\bibfnamefont#1{#1}\fi
\expandafter\ifx\csname citenamefont\endcsname\relax
  \def\citenamefont#1{#1}\fi
\expandafter\ifx\csname url\endcsname\relax
  \def\url#1{\texttt{#1}}\fi
\expandafter\ifx\csname urlprefix\endcsname\relax\def\urlprefix{URL }\fi
\providecommand{\bibinfo}[2]{#2}
\providecommand{\eprint}[2][]{\url{#2}}

\bibitem[{\citenamefont{Greiner et~al.}(2002)\citenamefont{Greiner, Mandel,
  Esslinger, Hansch, and Bloch}}]{GreinerNature}
\bibinfo{author}{\bibfnamefont{M.}~\bibnamefont{Greiner}},
  \bibinfo{author}{\bibfnamefont{O.}~\bibnamefont{Mandel}},
  \bibinfo{author}{\bibfnamefont{T.}~\bibnamefont{Esslinger}},
  \bibinfo{author}{\bibfnamefont{T.~W.} \bibnamefont{Hansch}},
  \bibnamefont{and} \bibinfo{author}{\bibfnamefont{I.}~\bibnamefont{Bloch}},
  \bibinfo{journal}{Nature} \textbf{\bibinfo{volume}{415}}, \bibinfo{pages}{39}
  (\bibinfo{year}{2002}).

\bibitem[{\citenamefont{Bloch et~al.}(2008)\citenamefont{Bloch, Dalibard, and
  Zwerger}}]{RevModPhys.80.885}
\bibinfo{author}{\bibfnamefont{I.}~\bibnamefont{Bloch}},
  \bibinfo{author}{\bibfnamefont{J.}~\bibnamefont{Dalibard}}, \bibnamefont{and}
  \bibinfo{author}{\bibfnamefont{W.}~\bibnamefont{Zwerger}},
  \bibinfo{journal}{Rev. Mod. Phys.} \textbf{\bibinfo{volume}{80}},
  \bibinfo{pages}{885} (\bibinfo{year}{2008}).

\bibitem[{\citenamefont{Fallani et~al.}(2007)\citenamefont{Fallani, Lye,
  Guarrera, Fort, and Inguscio}}]{PhysRevLett.98.130404}
\bibinfo{author}{\bibfnamefont{L.}~\bibnamefont{Fallani}},
  \bibinfo{author}{\bibfnamefont{J.~E.} \bibnamefont{Lye}},
  \bibinfo{author}{\bibfnamefont{V.}~\bibnamefont{Guarrera}},
  \bibinfo{author}{\bibfnamefont{C.}~\bibnamefont{Fort}}, \bibnamefont{and}
  \bibinfo{author}{\bibfnamefont{M.}~\bibnamefont{Inguscio}},
  \bibinfo{journal}{Phys. Rev. Lett.} \textbf{\bibinfo{volume}{98}},
  \bibinfo{pages}{130404} (\bibinfo{year}{2007}).

\bibitem[{\citenamefont{Pasienski et~al.}(2010)\citenamefont{Pasienski, McKay,
  White, and DeMarco}}]{DeMarco}
\bibinfo{author}{\bibfnamefont{M.}~\bibnamefont{Pasienski}},
  \bibinfo{author}{\bibfnamefont{D.}~\bibnamefont{McKay}},
  \bibinfo{author}{\bibfnamefont{M.}~\bibnamefont{White}}, \bibnamefont{and}
  \bibinfo{author}{\bibfnamefont{B.}~\bibnamefont{DeMarco}},
  \bibinfo{journal}{Nature Physics} \textbf{\bibinfo{volume}{6}},
  \bibinfo{pages}{677} (\bibinfo{year}{2010}).

\bibitem[{\citenamefont{Fisher et~al.}(1989)\citenamefont{Fisher, Weichman,
  Grinstein, and Fisher}}]{PhysRevB.40.546}
\bibinfo{author}{\bibfnamefont{M.~P.~A.} \bibnamefont{Fisher}},
  \bibinfo{author}{\bibfnamefont{P.~B.} \bibnamefont{Weichman}},
  \bibinfo{author}{\bibfnamefont{G.}~\bibnamefont{Grinstein}},
  \bibnamefont{and} \bibinfo{author}{\bibfnamefont{D.~S.}
  \bibnamefont{Fisher}}, \bibinfo{journal}{Phys. Rev. B}
  \textbf{\bibinfo{volume}{40}}, \bibinfo{pages}{546} (\bibinfo{year}{1989}).

\bibitem[{\citenamefont{Jaksch et~al.}(1998)\citenamefont{Jaksch, Bruder,
  Cirac, Gardiner, and Zoller}}]{PhysRevLett.81.3108}
\bibinfo{author}{\bibfnamefont{D.}~\bibnamefont{Jaksch}},
  \bibinfo{author}{\bibfnamefont{C.}~\bibnamefont{Bruder}},
  \bibinfo{author}{\bibfnamefont{J.~I.} \bibnamefont{Cirac}},
  \bibinfo{author}{\bibfnamefont{C.~W.} \bibnamefont{Gardiner}},
  \bibnamefont{and} \bibinfo{author}{\bibfnamefont{P.}~\bibnamefont{Zoller}},
  \bibinfo{journal}{Phys. Rev. Lett.} \textbf{\bibinfo{volume}{81}},
  \bibinfo{pages}{3108} (\bibinfo{year}{1998}).

\bibitem[{\citenamefont{Auerbach}(1994)}]{Auerbach}
\bibinfo{author}{\bibfnamefont{A.}~\bibnamefont{Auerbach}},
  \emph{\bibinfo{title}{Interacting Electrons and Quantum magnetism}}
  (\bibinfo{publisher}{Springer}, \bibinfo{address}{New York},
  \bibinfo{year}{1994}).

\bibitem[{\citenamefont{F\"olling et~al.}(2006)\citenamefont{F\"olling, Widera,
  M\"uller, Gerbier, and Bloch}}]{PhysRevLett.97.060403}
\bibinfo{author}{\bibfnamefont{S.}~\bibnamefont{F\"olling}},
  \bibinfo{author}{\bibfnamefont{A.}~\bibnamefont{Widera}},
  \bibinfo{author}{\bibfnamefont{T.}~\bibnamefont{M\"uller}},
  \bibinfo{author}{\bibfnamefont{F.}~\bibnamefont{Gerbier}}, \bibnamefont{and}
  \bibinfo{author}{\bibfnamefont{I.}~\bibnamefont{Bloch}},
  \bibinfo{journal}{Phys. Rev. Lett.} \textbf{\bibinfo{volume}{97}},
  \bibinfo{pages}{060403} (\bibinfo{year}{2006}).

\bibitem[{\citenamefont{Campbell et~al.}(2006)\citenamefont{Campbell, Mun,
  Boyd, Medley, Leanhardt, Marcassa, Pritchard, and Ketterle}}]{GretchenK}
\bibinfo{author}{\bibfnamefont{G.~K.} \bibnamefont{Campbell}},
  \bibinfo{author}{\bibfnamefont{J.}~\bibnamefont{Mun}},
  \bibinfo{author}{\bibfnamefont{M.}~\bibnamefont{Boyd}},
  \bibinfo{author}{\bibfnamefont{P.}~\bibnamefont{Medley}},
  \bibinfo{author}{\bibfnamefont{A.~E.} \bibnamefont{Leanhardt}},
  \bibinfo{author}{\bibfnamefont{L.~G.} \bibnamefont{Marcassa}},
  \bibinfo{author}{\bibfnamefont{D.~E.} \bibnamefont{Pritchard}},
  \bibnamefont{and} \bibinfo{author}{\bibfnamefont{W.}~\bibnamefont{Ketterle}},
  \bibinfo{journal}{Science} \textbf{\bibinfo{volume}{313}},
  \bibinfo{pages}{649} (\bibinfo{year}{2006}).

\bibitem[{\citenamefont{Rokhsar and Kotliar}(1991)}]{PhysRevB.44.10328}
\bibinfo{author}{\bibfnamefont{D.~S.} \bibnamefont{Rokhsar}} \bibnamefont{and}
  \bibinfo{author}{\bibfnamefont{B.~G.} \bibnamefont{Kotliar}},
  \bibinfo{journal}{Phys. Rev. B} \textbf{\bibinfo{volume}{44}},
  \bibinfo{pages}{10328} (\bibinfo{year}{1991}).

\bibitem[{\citenamefont{Krauth et~al.}(1992)\citenamefont{Krauth, Caffarel, and
  Bouchaud}}]{PhysRevB.45.3137}
\bibinfo{author}{\bibfnamefont{W.}~\bibnamefont{Krauth}},
  \bibinfo{author}{\bibfnamefont{M.}~\bibnamefont{Caffarel}}, \bibnamefont{and}
  \bibinfo{author}{\bibfnamefont{J.-P.} \bibnamefont{Bouchaud}},
  \bibinfo{journal}{Phys. Rev. B} \textbf{\bibinfo{volume}{45}},
  \bibinfo{pages}{3137} (\bibinfo{year}{1992}).

\bibitem[{\citenamefont{Freericks and Monien}(1996)}]{PhysRevB.53.2691}
\bibinfo{author}{\bibfnamefont{J.~K.} \bibnamefont{Freericks}}
  \bibnamefont{and} \bibinfo{author}{\bibfnamefont{H.}~\bibnamefont{Monien}},
  \bibinfo{journal}{Phys. Rev. B} \textbf{\bibinfo{volume}{53}},
  \bibinfo{pages}{2691} (\bibinfo{year}{1996}).

\bibitem[{\citenamefont{Elstner and Monien}(1999)}]{PhysRevB.59.12184}
\bibinfo{author}{\bibfnamefont{N.}~\bibnamefont{Elstner}} \bibnamefont{and}
  \bibinfo{author}{\bibfnamefont{H.}~\bibnamefont{Monien}},
  \bibinfo{journal}{Phys. Rev. B} \textbf{\bibinfo{volume}{59}},
  \bibinfo{pages}{12184} (\bibinfo{year}{1999}).

\bibitem[{\citenamefont{van Oosten et~al.}(2001)\citenamefont{van Oosten,
  van~der Straten, and Stoof}}]{PhysRevA.63.053601}
\bibinfo{author}{\bibfnamefont{D.}~\bibnamefont{van Oosten}},
  \bibinfo{author}{\bibfnamefont{P.}~\bibnamefont{van~der Straten}},
  \bibnamefont{and} \bibinfo{author}{\bibfnamefont{H.~T.~C.}
  \bibnamefont{Stoof}}, \bibinfo{journal}{Phys. Rev. A}
  \textbf{\bibinfo{volume}{63}}, \bibinfo{pages}{053601}
  (\bibinfo{year}{2001}).

\bibitem[{\citenamefont{Dickerscheid et~al.}(2003)\citenamefont{Dickerscheid,
  van Oosten, Denteneer, and Stoof}}]{PhysRevA.68.043623}
\bibinfo{author}{\bibfnamefont{D.~B.~M.} \bibnamefont{Dickerscheid}},
  \bibinfo{author}{\bibfnamefont{D.}~\bibnamefont{van Oosten}},
  \bibinfo{author}{\bibfnamefont{P.~J.~H.} \bibnamefont{Denteneer}},
  \bibnamefont{and} \bibinfo{author}{\bibfnamefont{H.~T.~C.}
  \bibnamefont{Stoof}}, \bibinfo{journal}{Phys. Rev. A}
  \textbf{\bibinfo{volume}{68}}, \bibinfo{pages}{043623}
  (\bibinfo{year}{2003}).

\bibitem[{\citenamefont{Schroll et~al.}(2004)\citenamefont{Schroll, Marquardt,
  and Bruder}}]{PhysRevA.70.053609}
\bibinfo{author}{\bibfnamefont{C.}~\bibnamefont{Schroll}},
  \bibinfo{author}{\bibfnamefont{F.}~\bibnamefont{Marquardt}},
  \bibnamefont{and} \bibinfo{author}{\bibfnamefont{C.}~\bibnamefont{Bruder}},
  \bibinfo{journal}{Phys. Rev. A} \textbf{\bibinfo{volume}{70}},
  \bibinfo{pages}{053609} (\bibinfo{year}{2004}).

\bibitem[{\citenamefont{Garc\'{i}a-Ripoll
  et~al.}(2004)\citenamefont{Garc\'{i}a-Ripoll, Cirac, Zoller, Kollath,
  Schollw\"{o}ck, and von Delft}}]{Garcia-Ripoll:04}
\bibinfo{author}{\bibfnamefont{J.~J.} \bibnamefont{Garc\'{i}a-Ripoll}},
  \bibinfo{author}{\bibfnamefont{J.}~\bibnamefont{Cirac}},
  \bibinfo{author}{\bibfnamefont{P.}~\bibnamefont{Zoller}},
  \bibinfo{author}{\bibfnamefont{C.}~\bibnamefont{Kollath}},
  \bibinfo{author}{\bibfnamefont{U.}~\bibnamefont{Schollw\"{o}ck}},
  \bibnamefont{and} \bibinfo{author}{\bibfnamefont{J.}~\bibnamefont{von
  Delft}}, \bibinfo{journal}{Opt. Express} \textbf{\bibinfo{volume}{12}},
  \bibinfo{pages}{42} (\bibinfo{year}{2004}).

\bibitem[{\citenamefont{Sengupta and Dupuis}(2005)}]{PhysRevA.71.033629}
\bibinfo{author}{\bibfnamefont{K.}~\bibnamefont{Sengupta}} \bibnamefont{and}
  \bibinfo{author}{\bibfnamefont{N.}~\bibnamefont{Dupuis}},
  \bibinfo{journal}{Phys. Rev. A} \textbf{\bibinfo{volume}{71}},
  \bibinfo{pages}{033629} (\bibinfo{year}{2005}).

\bibitem[{\citenamefont{Konabe et~al.}(2006)\citenamefont{Konabe, Nikuni, and
  Nakamura}}]{PhysRevA.73.033621}
\bibinfo{author}{\bibfnamefont{S.}~\bibnamefont{Konabe}},
  \bibinfo{author}{\bibfnamefont{T.}~\bibnamefont{Nikuni}}, \bibnamefont{and}
  \bibinfo{author}{\bibfnamefont{M.}~\bibnamefont{Nakamura}},
  \bibinfo{journal}{Phys. Rev. A} \textbf{\bibinfo{volume}{73}},
  \bibinfo{pages}{033621} (\bibinfo{year}{2006}).

\bibitem[{\citenamefont{Damski and Zakrzewski}(2006)}]{PhysRevA.74.043609}
\bibinfo{author}{\bibfnamefont{B.}~\bibnamefont{Damski}} \bibnamefont{and}
  \bibinfo{author}{\bibfnamefont{J.}~\bibnamefont{Zakrzewski}},
  \bibinfo{journal}{Phys. Rev. A} \textbf{\bibinfo{volume}{74}},
  \bibinfo{pages}{043609} (\bibinfo{year}{2006}).

\bibitem[{\citenamefont{Fialko et~al.}(2007)\citenamefont{Fialko, Moseley, and
  Ziegler}}]{PhysRevA.75.053616}
\bibinfo{author}{\bibfnamefont{O.}~\bibnamefont{Fialko}},
  \bibinfo{author}{\bibfnamefont{C.}~\bibnamefont{Moseley}}, \bibnamefont{and}
  \bibinfo{author}{\bibfnamefont{K.}~\bibnamefont{Ziegler}},
  \bibinfo{journal}{Phys. Rev. A} \textbf{\bibinfo{volume}{75}},
  \bibinfo{pages}{053616} (\bibinfo{year}{2007}).

\bibitem[{\citenamefont{dos Santos and Pelster}(2009)}]{PhysRevA.79.013614}
\bibinfo{author}{\bibfnamefont{F.~E.~A.} \bibnamefont{dos Santos}}
  \bibnamefont{and} \bibinfo{author}{\bibfnamefont{A.}~\bibnamefont{Pelster}},
  \bibinfo{journal}{Phys. Rev. A} \textbf{\bibinfo{volume}{79}},
  \bibinfo{pages}{013614} (\bibinfo{year}{2009}).

\bibitem[{\citenamefont{Danshita and Naidon}(2009)}]{PhysRevA.79.043601}
\bibinfo{author}{\bibfnamefont{I.}~\bibnamefont{Danshita}} \bibnamefont{and}
  \bibinfo{author}{\bibfnamefont{P.}~\bibnamefont{Naidon}},
  \bibinfo{journal}{Phys. Rev. A} \textbf{\bibinfo{volume}{79}},
  \bibinfo{pages}{043601} (\bibinfo{year}{2009}).

\bibitem[{\citenamefont{Polak and Kope\ifmmode~\acute{c}\else
  \'{c}\fi{}}(2007)}]{PhysRevB.76.094503}
\bibinfo{author}{\bibfnamefont{T.~P.} \bibnamefont{Polak}} \bibnamefont{and}
  \bibinfo{author}{\bibfnamefont{T.~K.}
  \bibnamefont{Kope\ifmmode~\acute{c}\else \'{c}\fi{}}},
  \bibinfo{journal}{Phys. Rev. B} \textbf{\bibinfo{volume}{76}},
  \bibinfo{pages}{094503} (\bibinfo{year}{2007}).

\bibitem[{\citenamefont{Byczuk and Vollhardt}(2008)}]{PhysRevB.77.235106}
\bibinfo{author}{\bibfnamefont{K.}~\bibnamefont{Byczuk}} \bibnamefont{and}
  \bibinfo{author}{\bibfnamefont{D.}~\bibnamefont{Vollhardt}},
  \bibinfo{journal}{Phys. Rev. B} \textbf{\bibinfo{volume}{77}},
  \bibinfo{pages}{235106} (\bibinfo{year}{2008}).

\bibitem[{\citenamefont{Huber et~al.}(2008)\citenamefont{Huber, Theiler,
  Altman, and Blatter}}]{PhysRevLett.100.050404}
\bibinfo{author}{\bibfnamefont{S.~D.} \bibnamefont{Huber}},
  \bibinfo{author}{\bibfnamefont{B.}~\bibnamefont{Theiler}},
  \bibinfo{author}{\bibfnamefont{E.}~\bibnamefont{Altman}}, \bibnamefont{and}
  \bibinfo{author}{\bibfnamefont{G.}~\bibnamefont{Blatter}},
  \bibinfo{journal}{Phys. Rev. Lett.} \textbf{\bibinfo{volume}{100}},
  \bibinfo{pages}{050404} (\bibinfo{year}{2008}).

\bibitem[{\citenamefont{Sheshadri et~al.}(1993)\citenamefont{Sheshadri,
  Krishnamurthy, Pandit, and Ramakrishnan}}]{Sheshadri}
\bibinfo{author}{\bibfnamefont{K.}~\bibnamefont{Sheshadri}},
  \bibinfo{author}{\bibfnamefont{H.~R.} \bibnamefont{Krishnamurthy}},
  \bibinfo{author}{\bibfnamefont{R.}~\bibnamefont{Pandit}}, \bibnamefont{and}
  \bibinfo{author}{\bibfnamefont{T.~V.} \bibnamefont{Ramakrishnan}},
  \bibinfo{journal}{EPL (Europhysics Letters)} \textbf{\bibinfo{volume}{22}},
  \bibinfo{pages}{257} (\bibinfo{year}{1993}).

\bibitem[{\citenamefont{Negele and Orland}(1988)}]{NegeleOrland}
\bibinfo{author}{\bibfnamefont{J.~W.} \bibnamefont{Negele}} \bibnamefont{and}
  \bibinfo{author}{\bibfnamefont{H.}~\bibnamefont{Orland}},
  \emph{\bibinfo{title}{Quantum Many-particle Systems}}
  (\bibinfo{publisher}{Perseus Books}, \bibinfo{address}{New York},
  \bibinfo{year}{1988}).

\bibitem[{\citenamefont{Jackiw and Kerman}(1979)}]{JackiwKerman}
\bibinfo{author}{\bibfnamefont{R.}~\bibnamefont{Jackiw}} \bibnamefont{and}
  \bibinfo{author}{\bibfnamefont{A.}~\bibnamefont{Kerman}},
  \bibinfo{journal}{Physics Letters} \textbf{\bibinfo{volume}{71A}},
  \bibinfo{pages}{158} (\bibinfo{year}{1979}).

\bibitem[{\citenamefont{Berry}(1984)}]{MVBerry}
\bibinfo{author}{\bibfnamefont{M.~V.} \bibnamefont{Berry}},
  \bibinfo{journal}{Proceedings of the Royal Society of London. A. Mathematical
  and Physical Sciences} \textbf{\bibinfo{volume}{392}}, \bibinfo{pages}{45}
  (\bibinfo{year}{1984}).

\bibitem[{\citenamefont{Polini et~al.}(2005)\citenamefont{Polini, Fazio,
  MacDonald, and Tosi}}]{PhysRevLett.95.010401}
\bibinfo{author}{\bibfnamefont{M.}~\bibnamefont{Polini}},
  \bibinfo{author}{\bibfnamefont{R.}~\bibnamefont{Fazio}},
  \bibinfo{author}{\bibfnamefont{A.~H.} \bibnamefont{MacDonald}},
  \bibnamefont{and} \bibinfo{author}{\bibfnamefont{M.~P.} \bibnamefont{Tosi}},
  \bibinfo{journal}{Phys. Rev. Lett.} \textbf{\bibinfo{volume}{95}},
  \bibinfo{pages}{010401} (\bibinfo{year}{2005}).

\bibitem[{\citenamefont{Zulicke}(1988)}]{zulicke}
\bibinfo{author}{\bibfnamefont{U.}~\bibnamefont{Zulicke}}, \bibinfo{type}{{PhD}
  dissertation}, \bibinfo{school}{Indiana University},
  \bibinfo{address}{Department of Physics} (\bibinfo{year}{1988}).

\bibitem[{\citenamefont{Blaizot and Ripka}(1985)}]{BlaizotRipka}
\bibinfo{author}{\bibfnamefont{J.-P.} \bibnamefont{Blaizot}} \bibnamefont{and}
  \bibinfo{author}{\bibfnamefont{G.}~\bibnamefont{Ripka}},
  \emph{\bibinfo{title}{Quantum Theory of Finite Systems}}
  (\bibinfo{publisher}{The MIT Press}, \bibinfo{address}{Cambridge, Mass},
  \bibinfo{year}{1985}).

\bibitem[{\citenamefont{Tilahun}()}]{frustMF}
\bibinfo{author}{\bibfnamefont{D.}~\bibnamefont{Tilahun}},
  \bibinfo{note}{(unpublished)}.

\bibitem[{\citenamefont{Klauder}(1979)}]{PhysRevD.19.2349}
\bibinfo{author}{\bibfnamefont{J.~R.} \bibnamefont{Klauder}},
  \bibinfo{journal}{Phys. Rev. D} \textbf{\bibinfo{volume}{19}},
  \bibinfo{pages}{2349} (\bibinfo{year}{1979}).

\bibitem[{\citenamefont{Shankar}(1994)}]{Shankar}
\bibinfo{author}{\bibfnamefont{R.}~\bibnamefont{Shankar}},
  \emph{\bibinfo{title}{Principles of Quantum Mechanics}}
  (\bibinfo{publisher}{Springer}, \bibinfo{address}{New York},
  \bibinfo{year}{1994}).

\bibitem[{\citenamefont{Sachdev}(2000)}]{Sachdev}
\bibinfo{author}{\bibfnamefont{S.}~\bibnamefont{Sachdev}},
  \emph{\bibinfo{title}{Quantum Phase Transitions}}
  (\bibinfo{publisher}{Cambridge University Press},
  \bibinfo{address}{Cambridge, England}, \bibinfo{year}{2000}).

\bibitem[{\citenamefont{Sanchez-Palencia and Lewenstein}(2010)}]{disorder}
\bibinfo{author}{\bibfnamefont{L.}~\bibnamefont{Sanchez-Palencia}}
  \bibnamefont{and}
  \bibinfo{author}{\bibfnamefont{M.}~\bibnamefont{Lewenstein}},
  \bibinfo{journal}{Nature Physics} \textbf{\bibinfo{volume}{6}},
  \bibinfo{pages}{87} (\bibinfo{year}{2010}).

\bibitem[{\citenamefont{van Otterlo et~al.}(1995)\citenamefont{van Otterlo,
  Wagenblast, Baltin, Bruder, Fazio, and Sch\"on}}]{PhysRevB.52.16176}
\bibinfo{author}{\bibfnamefont{A.}~\bibnamefont{van Otterlo}},
  \bibinfo{author}{\bibfnamefont{K.-H.} \bibnamefont{Wagenblast}},
  \bibinfo{author}{\bibfnamefont{R.}~\bibnamefont{Baltin}},
  \bibinfo{author}{\bibfnamefont{C.}~\bibnamefont{Bruder}},
  \bibinfo{author}{\bibfnamefont{R.}~\bibnamefont{Fazio}}, \bibnamefont{and}
  \bibinfo{author}{\bibfnamefont{G.}~\bibnamefont{Sch\"on}},
  \bibinfo{journal}{Phys. Rev. B} \textbf{\bibinfo{volume}{52}},
  \bibinfo{pages}{16176} (\bibinfo{year}{1995}).

\end{thebibliography}
%%%%%%%%%%%%%%%%%%%%%%%%%%%%%%%%%%%%%%%%%%%%%%%%%%%%%%%%%%%%%%%%%%%%%%%

\end{document}